\documentclass[doublecol,figures]{epl2}

\usepackage{amsmath}
\usepackage{amsfonts}
\usepackage{amssymb}

\title{Mesoscale simulations of polymer dynamics in microchannel flows}
\shorttitle{Polymer dynamics in Poiseuille flow}
\author{L. Cannavacciuolo \and R. G. Winkler \and G. Gompper}
\shortauthor{L. Cannavacciuolo \etal}  \institute{Institut f\"ur
Festk\"orperforschung, Forschungszentrum J\"ulich, 52425 J\"ulich}
  \abstract{The
non-equilibrium structural and dynamical properties of flexible polymers
confined in a square microchannel and exposed to a Poiseuille flow are
investigated by mesoscale simulations. The chain length and the flow strength
are systematically varied. Two transport regimes are identified, corresponding
to weak and strong confinement. For strong confinement, the transport
properties are independent of polymer length. The analysis of the long-time
tumbling dynamics of short polymers yields non-periodic motion with a sublinear
dependence on the flow strength. We find distinct differences for
conformational as well as dynamical properties from results obtained for simple
shear flow. }
\begin{document}

\maketitle
\section{Introduction}

Confinement fundamentally alters the properties of dilute polymer solutions --
compared to bulk behavior -- when either the polymer radius of gyration is on
the order of the characteristic dimensions of its proximity
\cite{wall:78,broc:77,vlie:90,jend:03}, and/or an external field is applied,
e.g., a shear or pressure field.  In the first case, geometrical constraints
lead to a stretching of the polymer parallel to the surfaces
\cite{wall:78,broc:77,vlie:90,jend:03,jend:03_1,tege:04}. In the second case,
in addition to flow-induced deformations, polymer-surface hydrodynamic
interactions determine the polymer dynamics and leads to, e.g., cross-stream
migration \cite{agar:94,jend:04,usta:06,khar:06,stei:06,usta:07}. The migration
effect has been studied intensively for DNA-like molecules by computer
simulations and the relevance of hydrodynamic interactions has been confirmed.
 Experimental studies of individual DNA molecules in steady shear flow by
fluorescence microscopy have provided a wealth of information on single polymer
dynamics \cite{smith:99,ledu:99,schr:05,gera:06}. In particular, large
conformational changes have been revealed due to tumbling motion
\cite{schr:05,delg:06,wink:06_1}. Similar studies for flexible polymers
confined between surfaces or in narrow channels have not been conducted so far,
however, a similar complex dynamics can be expected.  Understanding of single
polymer behavior is of paramount importance for the emerging technology of
microfluidic devices. Insight into the detailed microscopic conformational,
dynamical, and transport properties of polymers, e.g., DNA, will help in the
conception and design of such devices. Moreover, such studies will contribute
to the understanding of the transport properties of biological macromolecules
through blood vessels.  The proper account of hydrodynamic interactions is
essential in simulation studies of fluid flows in channels as is emphasized by
the appearance of cross-streamline migration. Recently developed mesoscale
simulation techniques, such as Lattice Boltzmann simulations
\cite{ahlr:99,usta:07}, Brownian dynamics simulations with a hydrodynamic
tensor \cite{jend:03}, and multi-particle-collision dynamics
\cite{webs:05,nogu:05} (also called stochastic rotation dynamics), are well
suited for simulations of microchannels flows and are able to bridge the
length- and time-scale gap among the solvent and solute degrees of freedom
thereby taking hydrodynamic interactions adequately into account.  In this
letter, we will present results for the conformational and dynamical properties
of polymers confined in a square channel and exposed to a Poiseuille flow by
mesoscale computer simulations. Both, the polymer length as well as the
pressure gradient are systematically varied. The considered polymers fall into
the crossover regime from weak to strong confinement. Experiments
\cite{stei:06} and simulations \cite{jend:04} predict a distinct different flow
behavior in the two limits, which is confirmed by our simulations. The strong
confinement regime has received less attention so far, it is however most
relevant if microchannels are intended to be used for DNA characterization. In
addition, we discuss chain orientation and tumbling dynamics, aspects which
have not been addressed in previous studies.
\section{Simulation method, model}

We use a hybrid simulation approach to study the properties of flexible
polymers in flow, where molecular dynamics simulations (MD) are combined with
the multi-particle-collision dynamics (MPC) method for the solvent
\cite{male:99,ihle:01}. MPC is particle based and proceeds in two steps. In the
streaming step, the solvent  particles of mass $m$ move ballistically for a
time $h$. In the collision step, particles are sorted into the cells of a cubic
lattice of lattice constant $a$ and their relative velocities, with respect to
the center-of-mass velocity of each cell, are rotated around a random axis by
an angle $\alpha$. For every cell, mass, momentum, and energy are conserved.
The algorithm is described in detail in
refs.~\cite{male:99,kiku:03,tuez:03,ripo:04,ripo:05}. The fluid is confined in
a square channel of side length $L$ and periodic boundary conditions are
applied along the channel axis ($L_{\parallel}$). No-slip boundary conditions
are imposed by the bounce-back rule, as described in Ref. \cite{lamu:01}, and
flow is induced by a gravitational force ($F=mg$) acting on every fluid
particle.  The linear polymer is comprised of mass points of mass $M$, which
are connected by a strong harmonic potential with an equilibrium bond length
$b$. Excluded-volume interactions are taken into account by the well-known 12-6
truncated Lennard-Jones potential with the parameters $\sigma$ and $\epsilon$
\cite{muss:05}. Since we consider pressure-driven flows, no gravitational force
acts on the polymer.  The interaction of a polymer with the solvent is realized
by inclusion of its monomers in the MPC collision step \cite{male:00_1}.
Extensive studies of polymer dynamics confirm the validity of this procedure
\cite{male:00_1,ripo:04,muss:05,webs:05}.  Specifically, we employ the
parameters $\alpha = 130^\circ$, $h=0.1 \tau$, with $\tau =\sqrt{m a^2/k_B T}$
($k_B$ is Boltzmann's constant and $T$ the temperature), the number of
particles per cell $\rho =10$, $M=m\rho$, $b=\sigma=a$, the fluid mass density
$\varrho = \rho m/a^3$, $k_BT/\epsilon =1$, and the time step in MD simulations
$h_{MD}=5 \times 10^{-3} \tau$. Polymers with the monomer numbers $N=20$, $40$,
$80$, and $160$ are considered. To minimize self-interactions of periodic
images, the corresponding channel lengths are $L_{\parallel}= 40a$, $60a$,
$100a$, and $170a$, respectively.  Simulations of the pure solvent confirm that
the generated velocity profiles agree with the solution of the Navier-Stokes
equations for the considered geometry.
\begin{figure}
\includegraphics*[width=8.5cm]{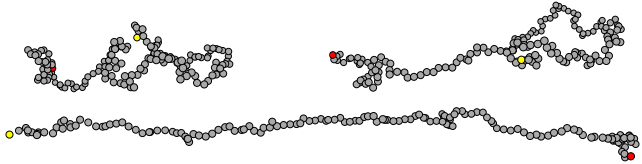} \caption{Conformations of a polymer
of length $N=160$ in a channel of width $L=15a$ for weak  (Peclet number
$Pe^L=10$, top two images) and strong flow ($Pe^L=100$, bottom). The latter
yields close to fully stretched polymers.
 } \label{fig_s}
\end{figure}

\begin{figure}
\onefigure[width=8cm]{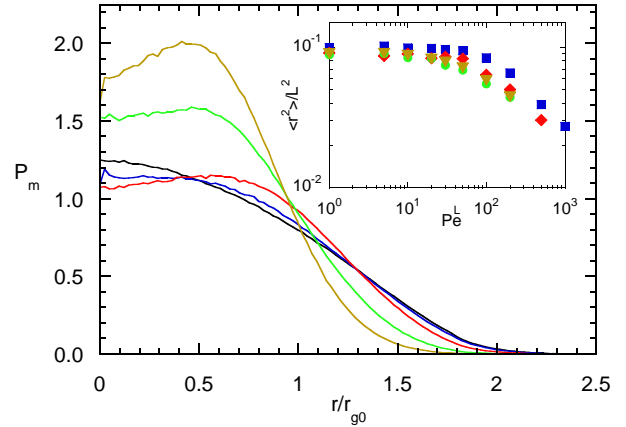} \caption{Radial monomer
distribution functions for the Peclet numbers $Pe^L = 0$, $10$, $50$, $100$,
$200$ (top to bottom at $r/r_{g0} =1.5 $), $N=40$, and $L/a=15$. Inset: widths
of the distribution functions for the chain lengths $N=20$ (squares), $N=40$
(diamonds), $N=80$ (bullets), and $N=160$ (triangles). } \label{fig_1}
\end{figure}

\section{Results}

We expect a pronounced different behavior of polymers in Poiseuille flow for
radii of gyration either smaller or larger than the width of the channel
\cite{stei:06}. Scaling considerations based on a blob model \cite{wall:78} and
simulations \cite{jend:03} yield an equilibrium stretch of a flexible polymer
in a microchannel when its bulk radius of gyration $r_{g0} \sim b N^{3/5}$
exceeds the lateral channel dimension, i.e., for $r_{g0} \gg L$. Here, the
channel width becomes the characteristic length scale and the blob relaxation
time, which is determined by $L$, will be the relevant time scale
\cite{jend:03,jend:04}. For $r_{g0} \ll L$, the conformations are unaffected by
confinement and the characteristic length and time scales are given by the bulk
values of the polymer. The ratios $2r_{g0}/L$ for  the considered polymer
lengths and the channel width $L=15a$ cover the range $0.36 - 1.3$,
corresponding to the crossover regime between weak and strong confinement. The
snapshots of fig.~\ref{fig_s} illustrate the preferred average alignment of a
polymer along the flow direction.
\begin{figure}
\onefigure[width=8cm]{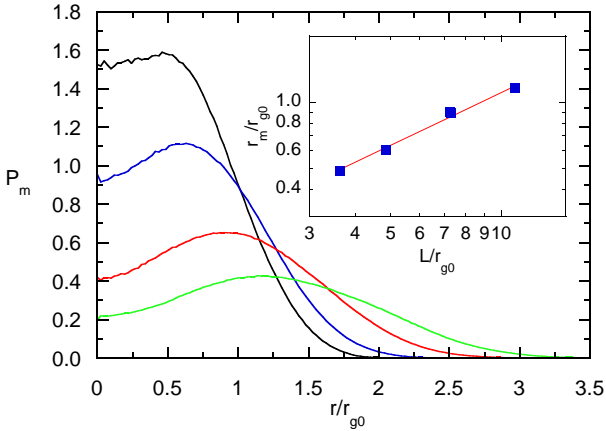} \caption{Radial monomer
distribution functions for the channel widths $L/a =15$, $20$, $30$, and $45$
(left to right). The polymer length is $N=40$ and $g \tau^2/a = 0.01$, i.e.,
the curvature of the flow profile is independent of $L$. Inset: position of the
maximum as function of the channel width. The straight line represents the
power law $(L/r_{g0})^{4/5}$. } \label{fig_2}
\end{figure}

\begin{figure}
\onefigure[width=8cm]{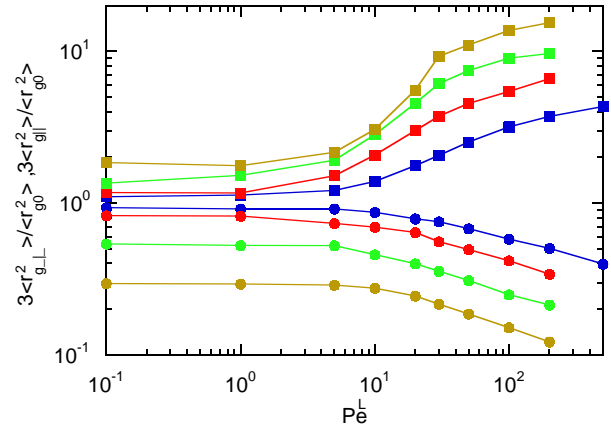} \caption{Mean square radii of gyration
parallel (squares) and transverse (bullets) to the channel for the chain
lengths $N=20$, $40$, $80$, $160$ (bottom to top, squares; top to bottom,
bullets) and $L=15a$ as a function of the Peclet number $Pe^L$ .} \label{fig_3}
\end{figure}

Figure~\ref{fig_1} provides an example of the radial monomer distribution for
the polymer length $N=40$ and various Peclet numbers. Here, we defined the
Peclet number $Pe^L$ as $Pe^L = \dot \gamma \tau_L$, with an effective shear
rate $\dot \gamma = g L \varrho /(2 \eta)$ and the relaxation time $\tau_L =
\eta (L/2)^3 /(3 k_B T)$ ($\eta$ is the solvent viscosity) \cite{doi:86}, where
the blob diameter has been replaced by the channel width. A similar approach
has been adopted in ref.~\cite{jend:04}. The distribution functions are
normalized such that $\int_0^{\infty} P_m(r/r_{g0}) r/r_{g0}^2 dr =1 $. With
increasing Peclet number, the distribution $P_m$ increases at $r/r_{g0} \approx
1$ due to the stretching of the polymer by the flow. For even large $Pe^L$,
migration sets in and the maximum of $P_m$ shifts to smaller radii, as already
observed in number of other studies
\cite{agar:94,jend:04,usta:06,khar:06,usta:07}. The widths of the radial
distribution functions
\begin{equation} \label{eq1}
\left\langle r^2 \right\rangle = \int_{0}^{\infty} r^3 P_m(r) dr
\end{equation}
for the various chain lengths are shown as inset in fig.~\ref{fig_1}. For small
Peclet numbers, the widths are independent of $Pe^L$ and decay at large $Pe^L$.
There seems to be a weak dependence on chain length only.  The widths start to
decrease at approximately the same  Peclet number. This is consistent with our
assumption that the blobs determine the polymer properties. The approach to a
universal limiting curve with increasing polymer length is explained as
follows. In general, $P_m$ exhibits the dependence $P_m (b,r,L,N,g)$, and
correspondingly the width $W^2 = \left\langle r^2 \right\rangle = W^2
(b,L,N,g)$. Scaling all lengths by $L$ and using the above definition of the
Peclet number yields the dependence $W^2/L^2 = f(r_{g0}/L,bN/L,Pe^L)$. For
strongly confined polymers, the width is determined by blobs -- and thus by $L$
-- rather than by the polymer length, hence, we find $W^2/L^2 = f(Pe^L)$, i.e.,
it depends on the Peclet number only.  According to current understanding,
migration is caused by polymer-surface hydrodynamic interactions
\cite{agar:94,jend:04,usta:06,khar:06,usta:07}. Since hydrodynamic interactions
are long-range in nature, it is difficult to unravel the system size dependence
of migration. To study the influence of the channel width on migration, we
performed simulations for various channel widths $L$ at the same curvature of
the flow profile ($g \tau^2/a = 0.01$) and polymer length $N=40$. As shown in
fig.~\ref{fig_2}, cross-streamline migration is well pronounced for all $L$,
with a shift of the maximum of the monomer distribution to larger radii with
increasing $L$. The position of the maximum ($r_m$) exhibits the chain-length
dependence $r_m \sim L^{4/5}$ as shown in the inset of fig.~\ref{fig_2}, i.e.,
migration is more pronounced for less strongly confined polymers.  The changes
in polymer conformations are reflected in their radii of gyration, which are
displayed in fig.~\ref{fig_3}. Evidently, short polymers ($N<40$) experience
only minor conformational restrictions at equilibrium, whereas longer polymers
are stretched along the channel and squeezed in the transverse direction.
Beyond a certain Peclet number $Pe^L_c$, flow induces or enhances stretching
parallel to the channel and transverse shrinkage. This $Pe^L_c$ value depends
only very weakly on chain length.

\begin{figure}
\onefigure[width=8cm]{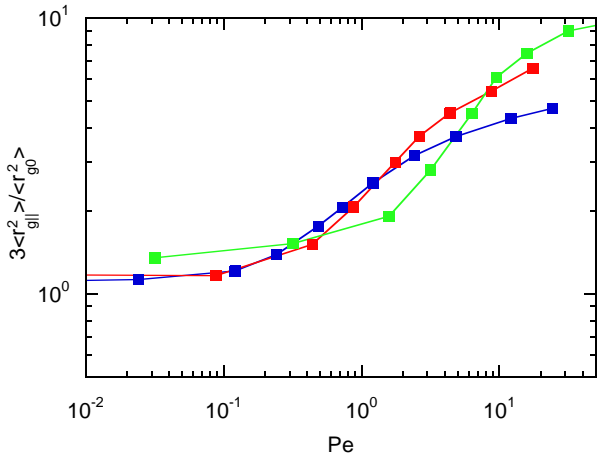} \caption{Mean square radii of
gyration parallel to the channel for the chain lengths $N=20$, $40$, and $80$
(bottom to top) and $L=15 a$ as a function of the Peclet number $Pe$. }
\label{fig_3_1}
\end{figure}

Figure~\ref{fig_3_1} shows the component of the mean square radius of gyration
along the channel axis for various chain lengths. Here, a different kind of
Peclet number is employed, which is defined as $Pe = \dot \gamma \tau_G$, with
$\tau_G = \eta r_{g0}^3/(3 k_B T)$  \cite{doi:86}, i.e., the polymer relaxation
time is used instead of a blob-related time. The Peclet number is then
chain-length dependent. The shorter polymers seem to exhibit universal behavior
for Peclet numbers up to $Pe \approx 5$. For larger $Pe$, finite polymer-size
effects play a role. However, the longer polymers show a different behavior.
This confirms our picture of the various involved length and time scales. The
polymers of lengths $N<40$ are only weakly confined in the channel. The flow
behavior is determined by the equilibrium radius of gyration and the longest
polymer relaxation time. Instead of the effective shear rate $\dot \gamma = g L
\varrho /(2 \eta)$, we could use a value defined in a different way, e.g., the
local shear rate averaged  over the channel width. For longer polymers ($N
> 40$), the channel width defines the relevant length and time scales as long
as the confinement is felt by the polymer. At high flow velocities,  the
polymer migrates to the channel center and is no longer in contact with the
surfaces (cf. fig.~\ref{fig_1}).

\begin{figure}
\onefigure[width=8cm]{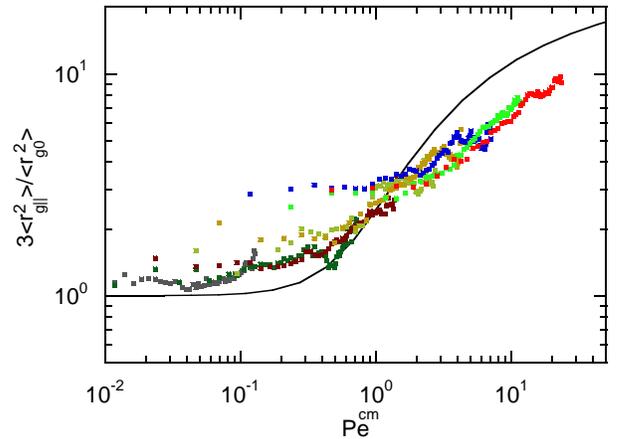} \caption{Mean square radii of
gyration parallel to the channel for the Peclet numbers $Pe^L =1$, $5$, $10$,
$20$, $30$, $50$, $100$, $200$ (left to right) as a function of the Peclet
number $Pe^{cm}$, where $Pe^{cm} \sim r_{cm}$. The solid line is an analytical
result for a chain of the same persistence length in simple shear flow
\cite{wink:06_1}. } \label{fig_4}
\end{figure}

So far, we have discussed conformational properties which are averaged over the
microchannel. Due to the parabolic velocity profile, however, the local shear
rate changes linearly with the radial distance. Thus, we can define a local
Peclet number by $Pe^{cm}= \varrho g r_{g0}^3 r_{cm} /(3 k_B T)$, where
$r_{cm}$ is the radial chain center-of-mass position. Figure~\ref{fig_4} shows
the dependence of the mean square radius of gyration parallel to the channel as
function of $Pe^{cm}$ for the chain length $N=40$ and $L=15a$. The various data
sets cover the range of Peclet numbers $Pe^L =1 - 200$. For small $Pe^{cm}$,
i.e., in the central part of the channel we find a stretching of the polymer by
the flow field which increases with increasing field strength $Pe^L$. Thus, in
the central part, the channel flow is different from a simple shear flow -- the
extended polymer never experience a linear velocity profile only. At larger
radial distances, the polymer is further stretched, particulary for larger
$Pe^L$ \cite{usta:06}. More importantly, the various data sets seem to approach
a limiting curve for sufficiently large $r_{cm}$. However, this curve deviates
from the theoretical dependence valid for linear shear flows, at least for the
length range covered by our simulations. The transverse radius of gyration
exhibits a similar deviation from the theoretical prediction. This leads us to
the conclusion that the non-equilibrium structural properties of flexible
polymers in a Poiseuille flow are even locally different from those of a
polymer in linear shear flow at least as long as the characteristic chain
dimensions are on the order of the channel width. The reason seems to be the
variation of the relevant relaxation times as $Pe^L$ and the center-of-mass
position change.

\begin{figure}
\onefigure[width=8cm]{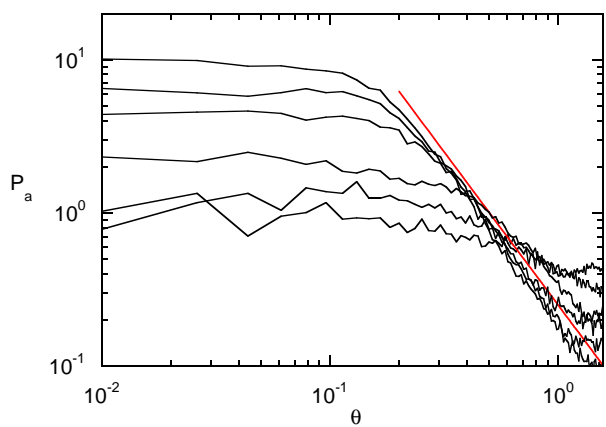} \caption{Probability distribution of
the angle between  the end-to-end vector and the flow direction for the Peclet
numbers $Pe^L =1$, $5$, $10$, $30$, $50$, $100$ (bottom to top) and the channel
width $L=15a$. Only half of a period is shown. The solid line represents the
power-law dependence $P_a \sim \theta^{-2}$. } \label{fig_5}
\end{figure}

\begin{figure}
\onefigure[width=8cm]{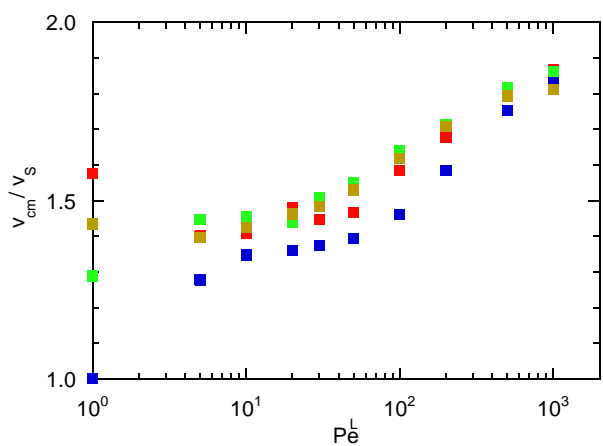} \caption{Ratio of the polymer
center-of-mass velocity $v_{cm}$ and the average solvent velocity $v_s$ for the
chain lengths $N=20$ (blue), $40$ (red), $80$ (green), and $160$ (orange) and
the channel width $L=15a$.} \label{fig_7}
\end{figure}

\begin{figure}
\onefigure[width=8cm]{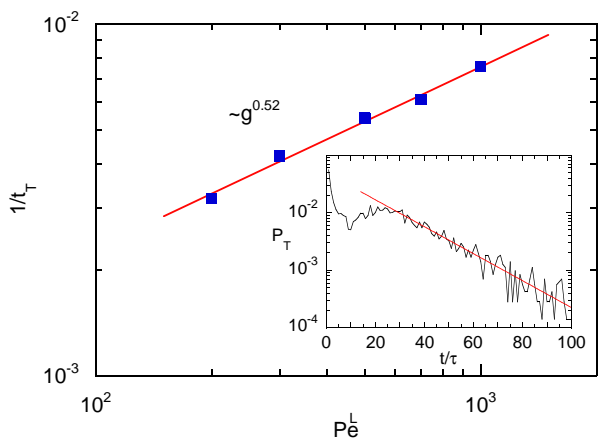} \caption{Tumbling times for the
chain length $N=20$ and the channel width $L=15a$. Inset: Distribution of
tumbling times for $Pe^L = 500$. } \label{fig_6}
\end{figure}

In addition, we analyze the orientational distribution function of the
end-to-end vector. Since the system is almost rotational symmetric with respect
to the channel axis, only the distribution function $P_a(\theta)$, where
$\theta$ ($0 < \theta < \pi$) is the angle between the end-to-end vector and
the flow direction, needs to be studied. As shown in fig.~\ref{fig_5}, the
distribution is almost independent of the angle $\theta$ at small Peclet
numbers. The deviation from the bulk value $1/2$ is caused by the confinement
of the polymer. (The distribution is normalized such that $\int_{0}^{\pi}
P_a(\theta) \sin \theta d \theta =1 $.) As expected, there is no preferred
angle; the average is $\left\langle \theta \right\rangle =0 $. The shape of the
distribution strongly depends on the Peclet number. For large Peclet numbers,
$P_a$ decays algebraically as $\theta^{-2}$. A similar algebraic decay is found
for the angle between the end-to-end vector and its projection onto the shear
plane of a flexible polymer in shear flow \cite{wink:06_1,gera:06}. However,
the distribution $P_a$ also reflects the fundamental difference of polymers in
Poiseuille and shear flow. For shear flow, there exists a regime of shear
rates, where the distribution is Gaussian, only for large shear rates follows
the algebraic decay \cite{gera:06,wink:06_1}. In fig.~\ref{fig_5} no Gaussian
regime is present. Because of the nonlinear flow profile, an algebraic decay is
not necessarily surprising, compared to simple shear, where this fact is
explained by higher order correlations and the importance of the whole time
history. Nevertheless, the similarity is striking and might be traced back to
the same origin.  The radial width of the monomer and center-of-mass
distributions decrease with increasing flow strength (cf. fig.~\ref{fig_2}) and
a polymer spends more time in the region of higher axial fluid velocity
\cite{jend:04}. This results in an increased polymer transport velocity
compared to the average solvent velocity, as shown in fig.~\ref{fig_7}. For a
wide channel $r_{g0} \ll L$, the ratio of $v_{cm}/v_s$ is close to unity at
equilibrium. Conformational restrictions for $r_{g0} \ll L$ lead to a narrowing
of the radial distribution functions \cite{stei:06} accompanied by increased
polymer flow velocities even at very weak imposed flows. This expected
transport behavior is shown in fig.~\ref{fig_7}. At small Peclet numbers $Pe^L
\lesssim 50$, we find -- except for $N=20$ -- center-of-mass velocities which
depend only weakly on $Pe^L$, caused by confinement. Only for the polymer of
length $N=20$ reduces the velocity ratio to unity at equilibrium. With
increasing Peclet number, the average polymer velocity increases, starting at
about the same Peclet number $Pe^L$ for all chain lengths, and the ratio
reaches almost its maximum value of two. We obtain a chain-length dependence of
the velocity ratio for short polymers only, which is rather a consequence of
the weaker confinement of the shorter polymer than a true molecular weight
dependence. The ratios for the longer polymers seem to approach a universal
limiting curve, which is consistent with the assumption that local properties
on a blob scale rather than global polymer properties determine their behavior.
As a consequence, no separation of highly confined DNA according to molecular
weight is possible.  Polymers in simple shear flow exhibit large conformational
changes due to tumbling motion, i.e., they stretch and recoil in the course of
time \cite{schr:05,gera:06,delg:06,wink:06_1}. Visual inspection of the polymer
dynamics in Poiseuille flow reveals similar large conformational fluctuations
(cf. fig.~\ref{fig_s}). Tumbling is characterized by a typical time denoted as
tumbling time $t_T$. There are various ways to extract such a characteristic
time, although the dynamics is non-periodic. We tried various strategies and
found the following approach most useful. We first determined the distribution
of times between consecutive crossings of the end-to-end vector with the plane
perpendicular to the flow direction, i.e., $\theta = \pm \pi/2$.  As shown in
the inset of fig.~\ref{fig_6}, this yields exponentially decaying distributions
($P_T \sim e^{-t/t_T}$) at large times. The extracted tumbling times $t_T$ are
presented in fig.~\ref{fig_6}. A power-law fit yields the dependence $t_T \sim
g^{-0.52}$. Simulations and analytical calculations
\cite{schr:05,gera:06,delg:06,wink:06_1} yield the dependence $t_T \sim  \dot
\gamma^{-2/3}$ for flexible polymers in linear shear flow. Hence, the tumbling
times of polymers in Poiseuille flow exhibit a weaker dependence on the
strength of the external field than in shear flow.

\section{Conclusions}

We have analyzed the flow behavior of flexible polymers confined in a square
channel with a side length comparable to the polymer radius of gyration. We
have identified two distinct transport regimes were polymers behave similarly.
For $r_{g0} \ll L$ the polymers are unaffected by the channel geometry and the
characteristic length and time scales are given by the radius of gyration and
the longest polymer relaxation time, respectively. In the strongly confined
regime $r_{g0} \gg L$, the polymer structure and dynamics are determined by
blobs with a size of the channel cross section. Polymers with $r_{g0} \approx
L/2$ show a complicated and non-universal behavior. We expect that the
properties of polymers in a slit geometry are less severely affected by
confinement than those of polymers in a channel. In the slit geometry, the
polymer is always able to relax in the direction parallel to the surfaces and
hence the longest relaxation time is always related to the polymer length. The
analysis of the radial polymer deformation, the distribution of the orientation
angle, and the tumbling times reveals differences among Poiseuille flow and
simple shear flow for the strongly confined polymers.


\end{document}